\documentclass{PoS}

\title{Lattice QCD Methods for Hadronic Polarizabilities}

\ShortTitle{Lattice QCD Methods for Hadronic Polarizabilities}

\author{
\speaker{B.~C.~Tiburzi}
\thanks{I am grateful to my insightful collaborators for fruitful collaboration on the various projects that are summarized in this talk. 
Work supported in part by a joint CCNY--RBRC fellowship, 
an award from the Professional Staff Congress of the CUNY, 
and by the U.S.~National Science Foundation, 
under Grant No.~PHY-$1205778$. 
}\\
        Department of Physics,
        The City College of New York,  
        New York, NY, USA\\
        Graduate School and University Center,
        The City University of New York,
        New York, NY, USA\\
        RIKEN BNL Research Center, 
        Brookhaven National Laboratory, 
        Upton, NY, USA\\
        E-mail: \email{btiburzi@ccny.cuny.edu}}

\abstract{
Chiral dynamics makes definitive predictions for the electromagnetic polarizabilities of hadrons near the chiral limit; 
but, 
agreement with experiment is tenuous in some cases.  
We provide an overview of lattice QCD methods to compute the electric and magnetic polarizabilities of hadrons. 
Central to these methods is the lattice simulation of quarks in uniform, classical electromagnetic fields.  
A long-term goal is the determination of polarizabilities directly from lattice computations, 
however, 
in the near term, 
one may need to rely on partially quenched chiral perturbation theory. 
Nonetheless the same striking predictions for the pion mass dependence of electric and magnetic polarizabilities can be made from chiral dynamics, 
and tested with lattice QCD. 
A particular focus is a novel new method to handle charged particle correlation functions in magnetic fields. 
}

\FullConference{The 7th International Workshop on Chiral Dynamics,\\
		August 6 -10, 2012\\
		Jefferson Lab, Newport News, Virginia, USA}

\begin{document}

\section{Electromagnetic Polarizabilities}

The electric and magnetic polarizabilities provide an opportunity for a stringent test of the chiral dynamics inside hadrons. 
Essential to the story of chiral dynamics is the spontaneous breakdown of chiral symmetry. 
For the case of two massless quark flavors, 
the QCD functional integral possesses an 
$SU(2)_L \otimes SU(2)_R$
symmetry, 
which is broken to the vector subgroup by the vacuum expectation value of the chiral condensate, 
$< \overline{\psi}_L \psi_R > \neq 0$. 
In this picture, 
the pions emerge as Goldstone modes, 
and have masses that vanish in the absence of explicit chiral symmetry breaking, 
that is
$m_\pi^2 = 0 + m_q  |< \overline \psi \psi > | F^2 $, 
where the explicit chiral symmetry breaking is parameterized by 
$m_q$,
the quark mass.
This picture is an effective description of low-energy QCD provided the quark masses are small compared to the scale of strong interactions. 
In terms of hadronic parameters that are free from QCD renormalization scale and scheme dependence, 
this condition translates into 
$m_\pi / (4 \pi F) \ll 1$; 
and, 
when met, 
allows for the quark mass dependence of low-energy QCD observables to determined systematically in an expansion about the chiral limit, 
$m_q = 0$. 

In chiral perturbation theory, 
the interactions of hadrons with pions are constrained by the form of spontaneous and explicit chiral symmetry breaking. 
As a result, 
the bare hadron fields are dressed with pions in a Fock state expansion, 
schematically of the form
\begin{eqnarray}
| \Pi^0 \rangle 
&=&
c_0 | \pi^0 \rangle + c_1 | \pi^0 \pi^+ \pi^- \rangle + \ldots, 
\qquad 
|N \rangle
=
c_0 | n \rangle + c_1 | p \pi^- \rangle + \ldots 
,\end{eqnarray}
where the bare fields are denoted with lower-case lettering, 
$c_0$'s 
are related to the wave-function renormalization, 
and 
$c_1$'s are determined by the interactions of the theory. 
Depending on kinematics, 
the higher Fock components may only be virtual states. 
Above we have chosen states that are electrically neutral to emphasize that dynamical fluctuations produce Fock components containing charged hadrons. 
The electromagnetic interaction thus serves as a probe sensitive to the higher Fock components of the bare hadron fields. 
Within chiral perturbation theory, 
the electric and magnetic polarizabilities
($\alpha_E$ and $\beta_M$, respectively) 
are determined to leading order entirely from higher Fock components. 
For the poins~\cite{Holstein:1990qy}, 
and nucleons~\cite{Bernard:1991rq}, 
there is a characteristic singularity in the chiral limit of these quantities, 
\begin{equation}
\alpha_E^{\pi}, 
\beta_M^\pi 
\sim 
\frac{1}{m_\pi} 
\sim 
\frac{1}{\sqrt{m_q}}, 
\qquad
\alpha_E^N, 
\beta_M^N
\sim 
\frac{1}{m_\pi} 
\sim 
\frac{1}{\sqrt{m_q}}
.\end{equation}
Experimental determination of polarizabilities can be achieved through the analysis of low-energy Compton scattering data. 
In principle, 
this is easiest for the proton, 
however, 
the low-energy limit is dominated by the total charge interaction, 
the Thomson cross section;
and, while increasing the energy leads to increased sensitivity to polarizabilities, 
it also introduces higher-order response of the nucleon. 
For the most recent comprehensive analysis of proton Compton scattering, see~\cite{McGovern:2012ew}.
For the neutron, 
one must use Compton scattering off deuterium or other light nuclei to extract polarizabilities, 
which introduces additional theoretical uncertainty. 
For pions, 
experiments have resorted to photo-pion production off the nucleon~\cite{Ahrens:2004mg}, 
and new results are anticipated from recent COMPASS measurements using pion scattering off Primakoff photons~\cite{Abbon:2007pq}.

\section{Methods for Particles in Electric Fields}

The electromagnetic polarizabilities provide an opportunity for lattice QCD computations, 
as they are quantities that have been subject to debate. 
A lattice determination of the polarizabilities of the deuteron, 
for example, 
would be a major contribution, 
as one would be able to study few-body dynamics in addition to the chiral behavior. 
Needless to say, 
lattice QCD is not yet at the point of such studies, 
however, 
there has been progress in treating strong interactions in the presence of uniform, 
classical electromagnetic fields, 
see~\cite{Tiburzi:2011vk} 
for overviews. 
One might think the natural starting point for lattice computations of polarizabilities would be the Compton scattering tensor, 
\begin{equation} \label{eq:T}
T_{\mu \nu} (k',k)
= 
\int d^4 x \, d^4 y \,
e^{ - i k'_\mu x_\mu + i k_\mu y_\mu}
\langle H | T\{ J_\mu(x) J_\nu(y) \} | H \rangle
,\end{equation}
however, 
this presents two major complications. 
The low-energy limit in 
Eq.~\ref{eq:T}
is constrained by spatial momentum quantization conditions resulting from periodic boundary conditions imposed on the quark fields.
Progress has been made on this front.  
The desired terms at second order in the photon frequency can be isolated by studying zero momentum derivatives of the quark propagators, 
and these can be computed approximately by using partially twisted boundary conditions, 
and taking the limit of vanishing twist angle~\cite{deDivitiis:2012vs}. 
This procedure leads one to the computation of certain hadronic four-point functions for which nearby intermediate states between the current insertions present an essential complication. 
While there has been progress in computing four-point functions in the case of the mesons, 
namely the
$K_L$--$K_S$ 
mass difference~\cite{Yu:2012nx}, 
it is likely that such methods will not be practicable for nucleons.

The external field approach provides an alternate method to access electromagnetic polarizabilities. 
One adds a classical electromagnetic field to QCD computations, 
and studies the subsequent external field dependence of hadronic correlation functions. 
Such dependence gives one access to hadronic couplings to the external field. 
To include an external electromagnetic field, 
one appends 
$U(1)$-valued links to the 
$SU(3)$ color gauge links
\begin{equation}
U_\mu(x)
\longrightarrow
U_\mu^{e.m.}(x)
U_\mu(x)
.\end{equation}
This must be done for valence quark propagators, 
and in the quark determinant used to generate gauge ensembles. 
The latter encompasses contributions due to the electric charges of sea quarks, 
and has only been achieved on lattices studying thermodynamics with staggered quarks. 
In the near term, 
calculations in weak external fields will exclude contributions from the sea quarks in gauge field generation. 
Gauge ensemble re-weighting techniques are a promising way to include sea quark charges%
~\cite{Freeman:2012cy}.
The light quark mass regime is problematic for the quenching of sea quark charges altogether, 
due to the so-called exceptionality of gauge configurations.
Exceptional configurations create an essential roadblock for post-multiplying electromagnetic links to 
existing gluon gauge configurations.  
Inasmuch as such configurations are not encountered, 
one can address the quenching of sea quark electric charges using chiral perturbation theory, 
for a discussion see%
~\cite{Tiburzi:2009yd}, 
and predictions exist for pion and nucleon polarizabilities as a function of the sea quark electric charges%
~\cite{Detmold:2006vu}. 
As our methods are addressed with exploratory lattice studies, 
we couple the external fields to valence quarks only.

A final word on the inclusion of uniform, external electromagnetic fields on a lattice. 
The periodicity of lattice quark fields leads to a quantization condition on the strength of external fields%
~\cite{'tHooft:1979uj}, 
because the hyper-torus forms a closed surface that does not leak any flux. 
For electric and magnetic fields, 
the quantization conditions are of the form%
\footnote{
Strictly speaking such quantization conditions do not lead to just uniform electromagnetic fields. 
There are also gauge-invariant, 
finite volume artifacts involving the non-trivial holonomy of the external field%
~\cite{Tiburzi:2008pa}. 
While such contributions should be exponentially suppressed, 
$\sim e^{ - m_\pi L}$, 
lattice results suggest that the effect might be non-negligible even at large pion masses%
~\cite{Engelhardt:2007ub}.
That study employs methods different than those outlined here.
}
\begin{equation}
q \mathcal{E} 
= 2 \pi n / L \beta, 
\qquad
q B  
= 2 \pi n / L^2
,\end{equation}
where
$L$ 
is the length of a spatial direction, with all three spatial directions assumed to be of the same length, 
and 
$\beta$ 
is the length of the temporal direction of the lattice. 
We write 
$\mathcal{E}$
for the electric field to make clear that we are in Euclidean space, 
where the action density would appear as
$\frac{1}{4} F_{\mu \nu} F_{\mu \nu} = \frac{1}{2} \left( \vec{B}^2 + \vec{\mathcal{E}}^2 \right)$.
Analytic continuation is necessary for results in Minkowski space;
but, 
as we are interested in quantities that are perturbative in the strength of the field, 
the continuation is trivial. 
The Schwinger mechanism%
~\cite{Schwinger:1951nm}, 
which is a non-perturbative phenomenon, 
is fortunately absent in Euclidean space. 
In this section, 
we concern ourselves with external electric fields.

\subsection{Neutral Particles}

For neutral hadrons, 
the method is simple to explain. 
One includes an external electric field, 
and measures two-point correlation functions of hadrons. 
In the long Euclidean time limit, 
these correlation functions should follow an exponential falloff, 
\begin{equation}
G_{\mathcal{E}} (\tau) 
= 
Z_{\mathcal{E}} \,
e^{- E(\mathcal{E}) \tau}, 
\end{equation}
where 
$E(\mathcal{E})$
is the energy of the hadron in the external electric field. 
For a neutron in an external electric field, 
the energy is given by%
~\cite{Detmold:2010ts}
\begin{equation}
E(\mathcal{E}) 
= 
M_n +  \left( \alpha_E - \mu_n^2 / 4 M_n^2 \right) \mathcal{E}^2 / 2 + \ldots
,\end{equation}
where terms of order 
$\mathcal{E}^4$
have been dropped. 
The contribution involving the square of the neutron magnetic moment arises from treating the neutron spin relativistically. 
In terms of neutron Compton scattering, 
the analogous contribution appears as a Born term, 
namely the second-order term arising from two interactions of leading-order. 
The exponential falloff of neutron correlators leads to an extraction of the quantity in parentheses. 
If one is interested in isolating the electric polarizability,
one requires a method to determine the neutron magnetic moment, 
and this can be achieved by looking at the amplitudes in off-diagonal spin components of the correlator%
~\cite{Detmold:2010ts}. 
In that study, 
an ensemble of anisotropic clover fermions~\cite{Edwards:2008ja}
was successfully used to demonstrate the technique. 
The analogue of the Born term was shown to affect extraction of the electric polarizability by 
$50 \%$.

\subsection{Charged Particles}

The study of charged particles in external electric fields is obviously complicated, 
however, 
the same philosophy can be applied. 
One determines hadronic correlation functions for various field strengths, 
and then matches onto the behavior expected from a single-particle effective action. 
This behavior is not a simple exponential falloff in Euclidean time. 
For example, 
the charged pion propagator should behave as%
~\cite{Tiburzi:2008ma}
\begin{equation}
G_\mathcal{E}(\tau)
=
Z_{\mathcal{E}}
\int_0^\infty \frac{ds}{\sqrt{\sinh Q \mathcal{E} s}}
e^{- \frac{1}{2} Q \mathcal{E} \tau^2 \coth Q \mathcal{E} s  - \frac{1}{2} E^2_\pi (\mathcal{E}) s}
,\end{equation}
with 
$E_\pi (\mathcal{E}) = m_\pi + \frac{1}{2} \alpha_E \mathcal{E}^2 + \cdots$.
While fits to the non-standard 
$\tau$-behavior 
can be challenging, 
the technique has been successfully demonstrated in an exploratory lattice calculation of charged pion and kaon correlation functions%
~\cite{Detmold:2009dx}.
Similar success was also achieved in the study of proton correlators using a generalization of the method to spin-half particles%
~\cite{Detmold:2010ts}.

\section{A Method for Charged Particles in Magnetic Fields}

The quantization condition for uniform magnetic fields has so far proven restrictive for the study of perturbatively small effects for hadrons. 
With increased lattice volumes, 
smaller magnetic field strengths can be accessed. 
The energy eigenstates of charged particles in external magnetic fields are described by Landau levels. 
For the charged pion, 
the long-time limit of the standard lattice correlation function should produce the standard exponential behavior
\begin{equation} \label{eq:B}
G_B(\tau)
= 
\sum_{\vec{x}}
\langle \pi(\vec{x}, \tau) \pi^\dagger(0,0) \rangle
= 
Z_B e^{ - E_0(B) \tau}
+ 
\cdots
,\end{equation}
where 
$E_0$
is the pion energy in the lowest Landau level, 
$E_0(B) = m_\pi + \frac{| Q B |}{ 2 m_\pi} + \frac{1}{2} \beta_M B^2 + \cdots$.
The omitted terms in the long-time limit of the correlation function include excited hadronic states, 
as well as the higher Landau levels. 
For large values of the external magnetic field, 
the Landau levels will be widely separated in energy, 
and only the lowest Landau level will survive the long-time limit. 
As we are interested in perturbatively small magnetic fields, 
however, 
the narrow Landau level spacing, 
$\Delta E / M = | Q B| / M^2$, 
will lead to a pileup.  
For smaller values of the magnetic field, 
one will require longer times to separate out the contribution from the lowest Landau level.

This complication can be sidestepped altogether%
~\cite{Tiburzi:2012ks}. 
Inspecting 
Eq.~\ref{eq:B}, 
we see that the sum over all lattice sites projects the correlator onto zero spatial momentum, 
$\vec{p} = 0$. 
In the presence of a magnetic field, 
it is impossible for all components of three-momentum to remain good quantum numbers. 
Consequently the correlator contains all Landau levels. 
A more judicious choice of correlation function is given by 
\begin{equation} \label{eq:Better}
\mathcal{G}_B(\tau)
= 
\sum_{\vec{x}}
\psi_0^* (x)
\langle \pi(\vec{x}, \tau) \pi^\dagger(0,0) \rangle
,\end{equation}
where 
$\psi_0(x)$
is the coordinate wave-function of the lowest Landau level. 
The long-time limit of the correlation function has the same exponential falloff, 
however, 
the first omitted terms are from higher lying hadronic states just as in the absence of the magnetic field. 
This can be demonstrated with the Schwinger proper-time trick. 
Explicit projection of the lowest Landau level should be an economic technique in perturbatively small magnetic fields.

As the technique must be practicable on the lattice, 
we investigated the effects of discretization on the lowest Landau level, 
and the effects of finite volume. 
For the former, 
we found that discretization effects on the wave-function of the lowest Landau level required in 
Eq.~\ref{eq:Better}
were negligible. 
The discretization corrections to the energy of the lowest Landau level could have a more substantial effect 
competitive with the magnetic polarizability, 
\begin{equation}
E^2_0 (B,a)
= 
m^2_\pi + | Q B | - 
\left( 
\frac{1}{8} a^2 Q^2 + m_\pi \beta_M
\right) B^2
.\end{equation}
Finite volume corrections where shown to be important; 
but, 
can be treated as an application of the magnetic periodicity of the action%
~\cite{AlHashimi:2008hr}.


\begin{thebibliography}{99}


\bibitem{Holstein:1990qy} 
  B.~R.~Holstein,
  Comments Nucl.\ Part.\ Phys.\ A {\bf 19}, 221 (1990).

\bibitem{Bernard:1991rq} 
  V.~Bernard, N.~Kaiser and U.~G.~Meissner,
  Phys.\ Rev.\ Lett.\  {\bf 67}, 1515 (1991);
  
  V.~Bernard, N.~Kaiser, A.~Schmidt and U.~G.~Meissner,
  Phys.\ Lett.\ B {\bf 319}, 269 (1993);
 
  T.~R.~Hemmert, B.~R.~Holstein and J.~Kambor,
  Phys.\ Rev.\ D {\bf 55}, 5598 (1997);

  S.~R.~Beane, M.~Malheiro, J.~McGovern, D.~Phillips and U.~van Kolck,
  Nucl.\ Phys.\ A {\bf 747}, 311 (2005).

\bibitem{McGovern:2012ew} 
  J.~A.~McGovern, D.~R.~Phillips and H.~W.~Griesshammer,
  arXiv:1210.4104 [nucl-th].

\bibitem{Ahrens:2004mg} 
  J.~Ahrens,  {\it et al.},
  Eur.\ Phys.\ J.\ A {\bf 23}, 113 (2005).

\bibitem{Abbon:2007pq} 
  P.~Abbon {\it et al.}  [COMPASS Collaboration],
  Nucl.\ Instrum.\ Meth.\ A {\bf 577}, 455 (2007).

\bibitem{Tiburzi:2011vk} 
  B.~C.~Tiburzi,
  PoS LATTICE {\bf 2011}, 020 (2011);
  
  M.~D'Elia,
  arXiv:1209.0374 [hep-lat].

\bibitem{deDivitiis:2012vs} 
  G.~M.~de Divitiis, R.~Petronzio and N.~Tantalo,
  Phys.\ Lett.\ B {\bf 718}, 589 (2012).

\bibitem{Yu:2012nx} 
  J.~Yu,
  PoS LATTICE {\bf 2012}, 129 (2012).

\bibitem{Freeman:2012cy} 
  W.~Freeman, A.~Alexandru, F.~Lee and M.~Lujan,
  PoS LATTICE {\bf 2012}, 015 (2012).

\bibitem{Tiburzi:2009yd} 
  B.~C.~Tiburzi,
  Phys.\ Rev.\ D {\bf 79}, 077501 (2009).
  
\bibitem{Detmold:2006vu} 
  W.~Detmold, B.~C.~Tiburzi and A.~Walker-Loud,
  Phys.\ Rev.\ D {\bf 73}, 114505 (2006);

  J.~Hu, F.~-J.~Jiang and B.~C.~Tiburzi,
  Phys.\ Rev.\ D {\bf 77}, 014502 (2008).

\bibitem{'tHooft:1979uj} 
  G.~'t Hooft,
  Nucl.\ Phys.\ B {\bf 153}, 141 (1979);

   J.~Smit and J.~C.~Vink,
  Nucl.\ Phys.\ B {\bf 286}, 485 (1987);
  
  P.~H.~Damgaard and U.~M.~Heller,
  Nucl.\ Phys.\ B {\bf 309}, 625 (1988).

\bibitem{Tiburzi:2008pa} 
  B.~C.~Tiburzi,
  Phys.\ Lett.\ B {\bf 674}, 336 (2009);

  W.~Detmold, B.~C.~Tiburzi and A.~Walker-Loud,
  arXiv:0908.3626 [hep-lat].

\bibitem{Engelhardt:2007ub} 
  M.~Engelhardt [LHPC Collaboration],
  Phys.\ Rev.\ D {\bf 76}, 114502 (2007).

\bibitem{Schwinger:1951nm} 
  J.~S.~Schwinger,
  Phys.\ Rev.\  {\bf 82}, 664 (1951).



\bibitem{Detmold:2010ts} 
  W.~Detmold, B.~C.~Tiburzi and A.~Walker-Loud,
  Phys.\ Rev.\ D {\bf 81}, 054502 (2010).

\bibitem{Edwards:2008ja} 
  R.~G.~Edwards, B.~Joo and H.~-W.~Lin,
  Phys.\ Rev.\ D {\bf 78}, 054501 (2008);

  H.~-W.~Lin {\it et al.}  [Hadron Spectrum Collaboration],
  Phys.\ Rev.\ D {\bf 79}, 034502 (2009).

\bibitem{Tiburzi:2008ma} 
  B.~C.~Tiburzi,
  Nucl.\ Phys.\ A {\bf 814}, 74 (2008).
  
\bibitem{Detmold:2009dx} 
  W.~Detmold, B.~C.~Tiburzi and A.~Walker-Loud,
  Phys.\ Rev.\ D {\bf 79}, 094505 (2009).



\bibitem{Tiburzi:2012ks} 
  B.~C.~Tiburzi and S.~O.~Vayl,
  arXiv:1210.4464 [hep-lat].

\bibitem{AlHashimi:2008hr} 
  M.~H.~Al-Hashimi and U.~-J.~Wiese,
  Annals Phys.\  {\bf 324}, 343 (2009).

\end{thebibliography}
\end{document}